\documentclass[onecolumn,showpacs,preprintnumbers,amsmath,amssymb,APSl,prd,superscriptaddress,nofootinbib]{revtex4-1}

\usepackage{dcolumn}% Align table columns on decimal point
\usepackage{bm}% bold math
\usepackage{ifpdf}
\usepackage{hyperref}
\usepackage{dcolumn}
\usepackage{bm}
\usepackage[spanish,english]{babel}
\usepackage{amsfonts}
\usepackage{amssymb}
\usepackage{graphicx}
\usepackage[latin1]{inputenc}

\newcommand{\be}{\begin{equation}}
\newcommand{\ee}{\end{equation}}
\newcommand{\bea}{\begin{eqnarray}}
\newcommand{\eea}{\end{eqnarray}}

\begin{document}

\title{Importance of torsion and invariant volumes in Palatini theories of gravity}

\author{Gonzalo J. Olmo} \email{ gonzalo.olmo@csic.es}
\affiliation{Departamento de F\'{i}sica Te\'{o}rica and IFIC, Centro Mixto Universidad de
Valencia - CSIC. Universidad de Valencia, Burjassot-46100, Valencia, Spain}
\author{D. Rubiera-Garcia} \email{drubiera@fisica.ufpb.br}
\affiliation{Departamento de F\'isica, Universidade Federal da
Para\'\i ba, 58051-900 Jo\~ao Pessoa, Para\'\i ba, Brazil}

\date{\today}

\begin{abstract}
We study the field equations of extensions of General Relativity formulated within a metric-affine formalism setting torsion to zero (Palatini approach). We find that different (second-order) dynamical equations arise depending on whether torsion is set to zero i) a priori or ii) a posteriori, i.e., before or after considering variations of the action.  Considering a generic family of Ricci-squared theories, we show that in both cases the connection can be decomposed as the sum of a Levi-Civita connection and terms depending on a vector field.  However, while in case i) this vector field is related to the symmetric part of the connection, in ii) it comes from the torsion part and,  therefore, it vanishes once torsion is completely removed. Moreover, the vanishing of this torsion-related vector field immediately implies the vanishing of the antisymmetric part of the Ricci tensor, which therefore plays no role in the dynamics. Related to this, we find that the Levi-Civita part of the connection is due to the existence of an invariant volume associated to an auxiliary metric $h_{\mu\nu}$,  which is algebraically related with the physical metric $g_{\mu\nu}$.
\end{abstract}

%\keywords{Modified gravity, Palatini approach, torsion}

%\arxivnumber{1306.4210 [hep-th]}

%Report: The article is well written and deserves a publication in PRD. I just suggest emphasizing the character of tensor density of many of the covariant derivatives appearing in the article; chapter 21 of Ref. [10] could be cited for this purpose. Also an explanation about the geometric character of l_\alpha, \sigma_\alpha, h_{\mu\nu}, etc. would facilitate the reading of the article. There is a typo in the definition of C_\mu.

\maketitle

\section{Introduction}

The relation between gravitation and geometry was established long ago by Einstein. He constructed his theory of general relativity (GR) using elements from Riemannian geometry and assuming that the metric tensor is enough to fully characterize the space-time geometry. His theory turned out to be in excellent agreement with observations and still remains valid today \cite{Will:2005va}. Nonetheless, there exist deep theoretical reasons to believe that the theory must be modified at high energies, where the quantum gravitational degrees of freedom are expected to become non-negligible \cite{beyondGR1,beyondGR2}. Observations of the large scale structure and dynamics of the universe \cite{supernovae1,supernovae2} have  been interpreted  as an indication that the infrared sector of the theory might also need some kind of modification \cite{infrared1,infrared2,infrared3,infrared4}. Einstein's theory could thus be seen as an approximation valid within a certain range of energies or length scales.

Phenomenological extensions of GR designed to address some of the above questions have followed different avenues, but most of them implicitly adopt the view that geometry is a matter of metrics. However, affine connections play a fundamental role in the implementation of general covariance and in the definition of curvature and torsion, which are tensorial quantities that provide an objective measure of the geometric properties of a manifold, regardless of the existence of a metric. Therefore, from a purely geometric perspective, one can consider a much wider framework in the important task of exploring the dynamics of geometric theories of gravity. Since there is no {\it a priori} reason, except tradition or convention,  to constrain the connection to be exactly defined by the Christoffel symbols of the metric at all energies and/or length scales, it seems natural to relax that constraint and explore its physical and technical implications.

Relaxing the metric constraints on the connection has allowed to make important progress in different directions. The first example was provided by Cartan, who found a fully covariant and geometric representation of Newtonian gravity in a purely affine context (without the need for a metric) \cite{MTW}. Cartan also considered the role of torsion, the antisymmetric part of the connection, in GR. His original idea was further developed by Sciama and Kibble, yielding a theory  in which the curvature of the metric is sourced by the energy-momentum of the matter sources whereas the torsion is sourced by their spin \cite{ECKS1,ECKS2}. The theory is experimentally as viable as GR \cite{Hammond} but has interesting new effects at high energies, being able to avoid the big bang singularity \cite{Hehl71,Hehl76,Poplawski11a,Poplawski11b}. Moreover, torsion effects may become important in particle physics \cite{Mavromatos:2012cc}
 and cosmology \cite{Fabri12,Fabri11,Belyaev:2007fn,Boehmer:2008ah} and in electroweak interactions \cite{Capo12}. In the context of extra dimensions a theory with torsion has been shown to be able to generate dynamical field equations in four dimensions different from those of GR \cite{Shankar11}. Torsion-based $f(T)$ gravities supporting black hole and brane solutions have been also examined in the literature \cite{Capo12b,Capo12bb, Ferraro:2011ks}. In a more fundamental string theory framework, it is well known that the Kalb-Ramond field strength acts like a torsion in many situations \cite{beyondGR1}, which may allow to experimentally probe some foundational aspects of the theory \cite{Hammond}. The effects of torsion have been explored in different string scenarios \cite{strings1,strings2,strings3,strings4}.

The exploration of the role played by connections needs not be restricted to the exploration of torsion effects. Symmetric connections are also relevant on their own. In particular, the original Hamiltonian formulation of GR by Arnowitt, Deser and Misner \cite{Arnowitt:1962hi}, was possible thanks to the use of the Palatini representation of the theory, where the metric and a symmetric connection are regarded as independent degrees of freedom. A related representation of classical GR with a symmetric connection, the Holst action \cite{Holst}, is also crucial for the non-perturbative quantization of the theory \cite{LQG1,LQG2,LQG3}.

Beyond GR, the Palatini approach \cite{Olmo:2011uz} has received increasing attention in the context of the cosmic speedup problem and in applications to quantum gravity phenomenology \cite{Olmo:2011sw}, the early universe, and black holes. It has been found, for instance, that $f(R)$ theories formulated \`{a} la Palatini naturally lead at late times to accelerating solutions. The reason is found in the fact that the field equations in vacuum are exactly equivalent to those of GR with a cosmological constant (GR+$\Lambda$), whose value depends on the particular $f(R)$ Lagrangian \cite{Olmo:2005hc}. This contrasts with the usual metric formulation of such theories \cite{Olmo:2006eh}, where a dynamical scalar degree of freedom is associated to the $f(R)$ function and, therefore, the solutions of the theory in vacuum are those of a scalar-tensor theory rather than GR+$\Lambda$.  On the other hand, the Palatini formulation of quadratic gravity, $R+a R^2+b R_{\mu\nu}R^{\mu\nu}$, for suitable choices of the parameters ($b>0$) also leads to non-singular solutions in the early universe, with the big bang replaced by a big bounce in isotropic and anisotropic scenarios \cite{Barragan:2010qb,Barragan:2009sq}. In that theory, charged black holes develop a non-trivial topological structure with the central singularity replaced by a wormhole, being certain configurations completely regular and possessing solitonic properties \cite{or12a1,or12a2,or12a3}, with interesting implications for the existence of black hole remnants and the dark matter problem \cite{Lobo:2013adx,Olmo:2013mla,Guendelman:2013sca}. In vacuum this Palatini theory is also equivalent to GR and no new dynamical degrees of freedom are present, which confirms the absence of ghosts and other instabilities that plague the standard metric formulation of the quadratic theory \cite{Zumino,Stelle} and is in agreement with the observed {\it universality of Einstein's equations} in Palatini theories of gravity \cite{Mauro1,Mauro2}. In addition, the field equations in the Palatini approach are always second-order, as opposed to the metric formulation of these theories.

Despite their appealing properties, Palatini theories have also been criticized in the literature on grounds of the Cauchy problem \cite{Sotiriou:2008rp,Far07} and on the existence of curvature divergences in the outermost regions of spherical systems with particular polytropic equations of state \cite{Poly1,Poly2} (see also \cite{PolyAlso1,PolyAlso2,PolyAlso3}). However, using a Hamiltonian approach \cite{Olmo:2011fh} it has been shown that the initial value problem is as well-formulated as in GR (because only up to second-order time derivatives appear) and that erroneous manipulations invalidate the conclusions of  \cite{Far07}.  The existence of surface singularities in spherical systems is also far from being a fundamental problem for these theories because it only affects to particular equations of state whose derivatives diverge as the pressure tends to zero. For sufficiently smooth equations of state, which are necessary to account for electrostatic effects and other phenomena not captured by the mere statistical properties of the matter distribution \cite{ST-1983}, the solutions are completely regular. Moreover, elementary objects with electric charge \cite{or12a1}, such as electrons or protons, do not exhibit such pathologies, which seem to be an artifact of the approximations  employed in the description of statistical/macroscopic systems.

Though much progress has been made in the Palatini approach and its applications in the last years, some basic aspects of this formulation are still poorly understood. In particular, in this work we consider the role of torsion in the derivation of the field equations and put forward the importance of invariant volumes in metric-affine theories to deal with the connection equations. To be precise, since in the Palatini approach one assumes that the connection is symmetric, i.e., it is torsionless, one may wonder if setting the torsion to zero {\it a priori} or {\it a posteriori} matters. In other words, are the field equations the same if i) one sets the torsion to zero before performing the variation of the action or if ii) it is set to zero after obtaining the equations? Focusing, in particular, on theories with Ricci-squared corrections, we will see that there is a difference, that assuming vanishing torsion {\it a priori} is, in general, not equivalent to setting it to zero {\it a posteriori}.  We will also show that it is possible to associate the symmetric part of the connection (or a part of it) with an invariant volume, which allows to obtain explicit (formal) solutions for the connection and facilitates the consideration of these theories in physical applications.

The paper is organized as follows: In section \ref{sec:II} we derive the field equations both for a symmetric connection and for connections with torsion. In section \ref{sec:III} we consider the particular case of Ricci-squared theories. First we assume the connection to be symmetric \emph{a priori}, and obtain and discuss the field equations, finding that the symmetric and antisymmetric parts of the Ricci tensor are directly coupled to each other.  Next we consider the presence of torsion and show that the symmetric and antisymmetric objects are not directly coupled, being (a part of) the torsion tensor the field that mediates their interaction. In both cases, the connection equations can be partially solved introducing a metric $h_{\mu\nu}$, related to the physical metric $g_{\mu\nu}$  by elementary algebraic transformations, but while in the former a dynamical vector field arises, in the latter it is absent once torsion is completely removed. In section \ref{sec:IV} we study the conditions for the existence of a well defined volume invariant and we conclude in section \ref{sec:V} with some comments and future prospects.

\section{Field equations} \label{sec:II}

In this section we derive the field equations corresponding to a generic Palatini theory in which the connection appears through the Riemann tensor suitably contracted with the metric and/or with itself.  We will first derive the field equations assuming that torsion is set to zero a priori, i.e., we assume a symmetric connection. Then we will consider the more general case, in which there are no a priori constraints on the connection, i.e., $\Gamma_{\beta \gamma}^{\alpha} \neq \Gamma_{\gamma \beta}^{\alpha}$ and, therefore, torsion is possible. The generic action that we consider  can be written as follows \cite{olmo11b,olmo11b2}
\begin{equation}\label{eq:f-action}
S=\frac{1}{2\kappa^2}\int d^4x \sqrt{-g}f(g_{\mu\nu},{R^\alpha}_{\beta\mu\nu})+S_m[g_{\mu\nu},\psi] \ ,
\end{equation}
where $f$ is some function of its arguments, $S_m$ is the matter action, $\psi$ represents collectively the matter fields, $\kappa^2$ is a constant with suitable dimensions (if $f=R$, then $\kappa^2=8\pi G$),  and the Riemann curvature tensor
is defined as follows
\begin{equation}\label{eq:Riemann} {R^\alpha}_{\beta\mu\nu}=\partial_\mu\Gamma_{\nu\beta}^\alpha-\partial_\nu\Gamma_{\mu\beta}^\alpha+\Gamma_{\mu\lambda}^\alpha\Gamma_{\nu\beta}^\lambda-\Gamma_{\nu\lambda}^\alpha\Gamma_{\mu\beta}^\lambda.
\end{equation}
From this definition, it is manifest that the only symmetry of the curvature tensor is ${R^\alpha}_{\beta\mu\nu}=-{R^\alpha}_{\beta\nu\mu}$, i.e., it is antisymmetric in its last two indices. We  assume a symmetric metric tensor $g_{\mu\nu}=g_{\nu\mu}$ and the usual definitions for the Ricci tensor $R_{\mu\nu}\equiv{R^\rho}_{\mu\rho\nu}$ and the Ricci scalar $R\equiv g^{\mu\nu}R_{\mu\nu}$.
Note that since we are working in the Palatini formalism,  no a priori relation between the metric and the connection is assumed. Thus the variation of the action (\ref{eq:f-action}) with respect to the metric and the connection can be expressed as
\begin{eqnarray}\label{eq:var1-f}
\delta S&=&\frac{1}{2\kappa^2}\int d^4x \sqrt{-g}\left[\left(\frac{\partial f}{\partial g^{\mu\nu}} -\frac{f}{2}g_{\mu\nu} \right)\delta g^{\mu\nu} + {P_\alpha}^{\beta\mu\nu}\delta {R^\alpha}_{\beta\mu\nu}\right]+\delta S_m \ ,
\end{eqnarray}
where we have used the notation  ${P_\alpha}^{\beta\mu\nu}\equiv \frac{\partial f}{\partial {R^\alpha}_{\beta\mu\nu}}$. The general form of  $\delta {R^\alpha}_{\beta\mu\nu}$ is given by
\begin{equation}
\delta {R^\alpha}_{\beta\mu\nu}= \nabla_\mu \left(\delta \Gamma^\alpha_{\nu\beta}\right)-\nabla_\nu \left(\delta\Gamma^\alpha_{\mu\beta}\right)+2S^\lambda_{\mu\nu}\delta\Gamma^\alpha_{\lambda\beta} \ ,
\end{equation}
 where $S^\lambda_{\mu\nu}\equiv ( \Gamma^\lambda_{\mu\nu}-\Gamma^\lambda_{\nu\mu})/2$ represents the torsion tensor, the antisymmetric part of the connection. It is at this point where one must decide whether to consider a purely symmetric  connection or if a torsion part is allowed.  In the following subsections we consider these two cases separately.

\subsection{Symmetric connections.}

In general, a connection  $\Gamma^\lambda_{\mu\nu}$ can be decomposed into its symmetric and antisymmetric parts as $\Gamma^\lambda_{\mu\nu}=C^\lambda_{\mu\nu}+S^\lambda_{\mu\nu}$, where $C^\lambda_{\mu\nu}=C^\lambda_{\nu\mu}$. Assuming that $S^\lambda_{\mu\nu}$ is zero {\it a priori},  the variation of the Riemann tensor becomes
\begin{equation}
\delta {R^\alpha}_{\beta\mu\nu}= \nabla_\mu \left(\delta \Gamma^\alpha_{\nu\beta}\right)-\nabla_\nu \left(\delta\Gamma^\alpha_{\mu\beta}\right) \ ,
\end{equation}
where $\delta \Gamma^\alpha_{\nu\beta}=\delta C^\alpha_{\nu\beta}$ and $\nabla_\mu\equiv \nabla_\mu^C$, i.e., it is the covariant derivative associated to the symmetric connection $C^\alpha_{\nu\beta}$.
In order to put the $\delta {R^\alpha}_{\beta\mu\nu}$ term in (\ref{eq:var1-f}) in suitable form, we need to note that
\begin{equation}\label{eq:step1s}
I_\Gamma=\int d^4x \sqrt{-g} {P_\alpha}^{\beta\mu\nu}\nabla_\mu \delta \Gamma^\alpha_{\nu\beta}=\int d^4x \left[\nabla_\mu(\sqrt{-g}J^\mu)-\delta \Gamma^\alpha_{\nu\beta}\nabla_\mu\left(\sqrt{-g} {P_\alpha}^{\beta\mu\nu}\right)\right] \ ,
\end{equation}
where $J^\mu\equiv {P_\alpha}^{\beta\mu\nu}\delta \Gamma^\alpha_{\nu\beta}$. As is well known (see \cite{MTW}, chapter 21, and \cite{Schouten}), the covariant derivative of a tensor density is, in general, given by $\nabla_\mu\sqrt{-g}=\partial_\mu \sqrt{-g}-\Gamma^\lambda_{\mu\lambda} \sqrt{-g}$, from which we find that       $\nabla_\mu(\sqrt{-g}J^\mu)=\partial_\mu(\sqrt{-g}J^\mu)$. With this result,  (\ref{eq:step1s}) can be cast as
\begin{equation}\label{eq:step2s}
I_\Gamma=\int d^4x \left[\partial_\mu(\sqrt{-g}J^\mu)-\delta \Gamma^\alpha_{\nu\beta}\nabla_\mu\left(\sqrt{-g} {P_\alpha}^{\beta\mu\nu}\right)\right] \ .
\end{equation}
Using this, (\ref{eq:var1-f}) becomes
\begin{eqnarray}\label{eq:var2s-f}
\delta S&=&\frac{1}{2\kappa^2}\int d^4x \left[\sqrt{-g}\left(\frac{\partial f}{\partial g^{\mu\nu}} -\frac{f}{2}g_{\mu\nu} \right)\delta g^{\mu\nu}+\partial_\mu\left(\sqrt{-g}J^\mu\right) \right. \nonumber \\
&-& \left. 2\nabla_\mu \left(\sqrt{-g}{P_\alpha}^{\beta[\mu\nu]} \right)\delta \Gamma^\alpha_{\nu\beta}\right]+\delta S_m \ ,
\end{eqnarray}
where ${P_\alpha}^{\beta[\mu\nu]}=({P_\alpha}^{\beta\mu\nu}-{P_\alpha}^{\beta\nu\mu})/2$.
Now, to make explicit the fact that  the metric and the connection are symmetric, this last expression can be rewritten as
\begin{eqnarray}\label{eq:var2salt-f}
\delta S&=&\frac{1}{2\kappa^2}\int d^4x \left[\sqrt{-g}\left(\frac{\partial f}{\partial g^{(\mu\nu)}} -\frac{f}{2}g_{\mu\nu} \right)\delta g^{\mu\nu}+\partial_\mu\left(\sqrt{-g}J^\mu\right) \right. \nonumber \\
&-& \left. \nabla_\mu \left[\sqrt{-g}\left({P_\alpha}^{\beta[\mu\nu]}+{P_\alpha}^{\nu[\mu\beta]} \right)\right]\delta C^\alpha_{\nu\beta}\right]+\delta S_m \ .
\end{eqnarray}
We thus find that the field equations can be written as follows
\begin{eqnarray}\label{eq:gmn_s}
\kappa^2 T_{\mu\nu}&=&\frac{\partial f}{\partial g^{(\mu\nu)}} -\frac{f}{2}g_{\mu\nu}  \\
\kappa^2{H_\alpha}^{\nu\beta}&=&-\frac{1}{\sqrt{-g}} \nabla_\mu^C \left[\sqrt{-g}\left({P_\alpha}^{\beta[\mu\nu]}+{P_\alpha}^{\nu[\mu\beta]} \right)\right] \ , \label{eq:Gamn_s}
\end{eqnarray}
where  $T_{\mu\nu}\equiv-\frac{2}{\sqrt{-g}}\frac{\delta S_m}{\delta g^{\mu\nu}}$ is the energy-momentum tensor of the matter, and $ {H_\alpha}^{\nu\beta}\equiv-\frac{1}{\sqrt{-g}}\frac{\delta S_m}{\delta \Gamma^\alpha_{\nu\beta}}$ represents the coupling of matter to the connection. For simplicity,   from now on we will assume that  ${H_\alpha}^{\nu\beta}=0$.

\subsection{Connections with torsion}

When the connection is not symmetric, one finds that $\nabla_\mu(\sqrt{-g}J^\mu)=\partial_\mu(\sqrt{-g}J^\mu)+2S^\sigma_{\sigma \mu}\sqrt{-g}J^\mu$. This implies that (\ref{eq:step1s}) turns into
\begin{equation}\label{eq:step2a}
I_\Gamma=\int d^4x \left[\partial_\mu(\sqrt{-g}J^\mu)-\delta \Gamma^\alpha_{\nu\beta}\left\{\nabla_\mu\left(\sqrt{-g} {P_\alpha}^{\beta\mu\nu}\right)-2S^\sigma_{\sigma \mu}\sqrt{-g}{P_\alpha}^{\beta\mu\nu}\right\}\right] \ ,
\end{equation}
which, as compared to (\ref{eq:step2s}), picks up a new term in the torsion $S^\sigma_{\sigma \mu}$. Using this result, (\ref{eq:var1-f}) becomes
\begin{eqnarray}\label{eq:var2-f}
\delta S&=&\frac{1}{2\kappa^2}\int d^4x \left[\sqrt{-g}\left(\frac{\partial f}{\partial g^{\mu\nu}} -\frac{f}{2}g_{\mu\nu} \right)\delta g^{\mu\nu}+\partial_\mu\left(\sqrt{-g}J^\mu\right) \right. \\
&+& \left.\left\{-\frac{1}{\sqrt{-g}}\nabla_\mu \left(\sqrt{-g}{P_\alpha}^{\beta[\mu\nu]} \right)+S^\nu_{\sigma\rho}{P_\alpha}^{\beta\sigma\rho}+2S^\sigma_{\sigma\mu}{P_\alpha}^{\beta[\mu\nu]}\right\}2\sqrt{-g}\delta \Gamma^\alpha_{\nu\beta}\right] \nonumber \\ &+& \delta S_m. \nonumber
\end{eqnarray}
Since the variations $\delta \Gamma^\alpha_{\nu\beta}$ and $\delta \Gamma^\alpha_{\beta\nu}$ are now independent, we thus find that the field equations can be written as follows
\begin{eqnarray}\label{eq:Gamn}
\kappa^2 T_{\mu\nu}&=&\frac{\partial f}{\partial g^{(\mu\nu)}} -\frac{f}{2}g_{\mu\nu}  \\
\kappa^2{H_\alpha}^{\nu\beta}&=&-\frac{1}{\sqrt{-g}}\nabla_\mu \left(\sqrt{-g}{P_\alpha}^{\beta[\mu\nu]} \right)+S^\nu_{\sigma\rho}{P_\alpha}^{\beta\sigma\rho}+2S^\sigma_{\sigma\mu}{P_\alpha}^{\beta[\mu\nu]} \ , \label{eq:Gamn_a}
\end{eqnarray}
where from now on we assume also that ${H_\alpha}^{\nu\beta}=0$. In order to compare (\ref{eq:Gamn_s}) with (\ref{eq:Gamn_a}) it is convenient to split the  connection into its symmetric and antisymmetric parts\footnote{Note that even though the connection is not a tensor, the difference between any two connections is a tensor, so we shall refer to $S^\alpha_{\mu\nu}$ as the torsion tensor.}, leading to $\nabla_\mu A_\nu=\partial_\mu A_\nu-C^\alpha_{\mu\nu} A_\alpha-S^\alpha_{\mu\nu} A_\alpha=\nabla_\mu^C A_\nu-S^\alpha_{\mu\nu} A_\alpha$ and $\nabla_\mu \sqrt{-g}=\nabla_\mu^C \sqrt{-g}-S^\alpha_{\mu\alpha}\sqrt{-g}$. By doing this, (\ref{eq:Gamn_a}) becomes
\begin{equation}
\kappa^2{H_\alpha}^{\nu\beta}=-\frac{1}{\sqrt{-g}}\nabla_\mu^C \left(\sqrt{-g}{P_\alpha}^{\beta[\mu\nu]} \right)+S^\lambda_{\mu\alpha}{P_\lambda}^{\beta[\mu\nu]}-S^\beta_{\mu\lambda}{P_\alpha}^{\lambda[\mu\nu]} \ . \label{eq:Gamn_a2}
\end{equation}
Setting the torsion to zero in this expression and comparing with (\ref{eq:Gamn_s}), it is easy to see that the field equations in these torsionless theories are different in general. The equivalence is limited to those Lagrangians satisfying ${P_\alpha}^{\beta[\mu\nu]}={P_\alpha}^{\nu[\mu\beta]}$.  To see the effects of this difference in detail, we consider next some illustrative examples.

\section{Ricci-squared theories } \label{sec:III}
We now consider a particular family of theories in which the Lagrangian is defined using invariants constructed only with the metric and the Ricci tensor or some parts of it.  This will allow us to make contact with relevant literature and obtain some new results.

Using the decomposition $\Gamma^\alpha_{\mu\nu}=C^\alpha_{\mu\nu}+S^\alpha_{\mu\nu}$, the Riemann tensor can be expressed as
\begin{equation} \label{Riemann-splitted}
R^{\alpha}_{\beta \mu \nu}(\Gamma)=R^{\alpha}_{\beta \mu \nu} (C)+ \nabla_{\mu} S^{\alpha}_{\nu \beta}- \nabla_{\nu} S^{\alpha}_{\mu\beta} + S^{\alpha}_{\mu \lambda}S^{\lambda}_{\nu \beta} - S^{\alpha}_{\nu \lambda} S^{\lambda}_{\mu \beta}.
\end{equation}
From the definition (\ref{eq:Riemann}), we define the Ricci tensor as
\begin{equation}\label{eq:Ricci}
R_{\beta\nu}\equiv {R^\alpha}_{\beta\alpha\nu}=\partial_\alpha\Gamma_{\nu\beta}^\alpha-
\partial_\nu\Gamma_{\alpha\beta}^\alpha+\Gamma_{\alpha\lambda}^\alpha\Gamma_{\nu\beta}^\lambda-\Gamma_{\nu\lambda}^\alpha\Gamma_{\alpha\beta}^\lambda,
\end{equation}
which can be recast as $R_{\beta\nu}(\Gamma)=B_{\beta\nu}(C)+\Omega_{\beta\nu}(S)$, where \cite{Eisenhart1}
\begin{eqnarray}
B_{\beta\nu}&=& \partial_\alpha C_{\nu\beta}^\alpha-
\partial_\nu C_{\alpha\beta}^\alpha+C_{\alpha\lambda}^\alpha C_{\nu\beta}^\lambda-C_{\nu\lambda}^\alpha C_{\alpha\beta}^\lambda \\
\Omega_{\beta\nu} &=& \nabla^C_\lambda S^\lambda_{\nu\beta}-\nabla^C_\nu S^\lambda_{\lambda\beta}+S^\lambda_{\nu\beta}S^\kappa_{\kappa\lambda}-S^\lambda_{\kappa\beta}S^\kappa_{\nu\lambda}.
\end{eqnarray}
From the above definitions one also finds that
\begin{equation} \label{Ricci-antisymmetric}
B_{[\beta \nu]}=\frac{1}{2} \left[ \nabla_{\beta} C^{\lambda}_{\lambda \nu} - \nabla_{\nu} C^{\lambda}_{\lambda \beta} \right] \ ,
\end{equation}
which indicates that the antisymmetric part of $B_{\beta\nu}$, which has been constructed with the symmetric part of the connection, is not zero in general (even in the torsionless case, where $B_{\beta\nu}$ coincides with $R_{\beta\nu}$).
 Also, the symmetric part of $\Omega_{\beta\nu} $ is not zero in general. Therefore, the most general action constructed only with the building blocks of the Ricci tensor and its contractions up to second order can be written as a functional of the scalars $g^{\mu\nu}B_{(\mu\nu)}$, $g^{\mu\nu}\Omega_{(\mu\nu)}$, $B^{(\mu\nu)}B_{(\mu\nu)}$, $\Omega^{(\mu\nu)}\Omega_{(\mu\nu)}$,  $B^{(\mu\nu)}\Omega_{(\mu\nu)}$, $B^{[\mu\nu]}B_{[\mu\nu]}$, $\Omega^{[\mu\nu]}\Omega_{[\mu\nu]}$,  $B^{[\mu\nu]}\Omega_{[\mu\nu]}$, and of the contractions of the dual tensors ${B^*}^{[\mu\nu]}{B^*}_{[\mu\nu]}$, ${\Omega^*}^{[\mu\nu]}{\Omega^*}_{[\mu\nu]}$,  ${B^*}^{[\mu\nu]}{\Omega^*}_{[\mu\nu]}$, where all the indices are raised with the metric.

We will first consider the field equations of theories constructed with some of the above invariants in the torsionless case (assuming that the torsion is zero {\it a priori}) and  then will consider their form when torsion is allowed.

\subsection{Spaces with a symmetric connection \label{sec:apriori}}

If the connection is assumed symmetric {\it a priori}, the objects that can appear in the action are just $g^{\mu\nu}B_{(\mu\nu)}$,  $B^{(\mu\nu)}B_{(\mu\nu)}$,  $B^{[\mu\nu]}B_{[\mu\nu]}$, and ${B^*}^{[\mu\nu]}{B^*}_{[\mu\nu]}$. We will refer to these theories generically as $f(B,Q_S,Q_A,Q^*_A)$, where $B\equiv g^{\mu\nu}B_{(\mu\nu)}$, $Q_S= B^{(\mu\nu)}B_{(\mu\nu)}$, $Q_A=B^{[\mu\nu]}B_{[\mu\nu]}$, and $Q^*_A={B^*}^{[\mu\nu]}{B^*}_{[\mu\nu]}$. Since in the literature only theories of the type $f(B,Q_S,Q_A)$ have been considered, for the moment we shall restrict our attention to this family (for simplicity, we will also assume that the matter is not coupled to the connection). In this case we find
\begin{equation}\label{eq:def_M}
{P_\alpha}^{\beta\mu\nu} = \delta_\alpha^\mu\left[f_B g^{\beta\nu}+2f_{Q_S} B^{(\beta\nu)}+2f_{Q_A} B^{[\beta\nu]}\right]\equiv \delta_\alpha^\mu M^{\beta\nu} \ ,
\end{equation}
which inserted in the field equations (\ref{eq:gmn_s}) and (\ref{eq:Gamn_s}) yields
\begin{eqnarray}
f_B B_{(\mu\nu)}+2g^{\alpha\beta}\left[f_{Q_S}B_{(\mu\alpha)}B_{(\nu\beta)}+f_{Q_A}B_{[\mu\alpha]}B_{[\nu\beta]}\right]-\frac{f}{2}g_{\mu\nu} & =& \kappa^2 T_{\mu\nu} \label{eq:met1s} \\
 \nabla^C_\mu\left[\sqrt{-g}\left(\delta_\alpha^\mu M^{(\beta\nu)}-\frac{1}{2}\left(\delta_\alpha^\nu M^{\beta\mu}+\delta_\alpha^\beta M^{\nu\mu}\right)\right)\right]  &=& 0 \ . \label{eq:con1s}
\end{eqnarray}
Since $M^{\beta\mu}=M^{(\beta\mu)}+M^{[\beta\mu]}$,  tracing over the indices $\alpha$ and $\nu$ in  (\ref{eq:con1s}), one finds that
\begin{equation}\label{eq:direct_coupling}
 \nabla_\lambda^C[\sqrt{-g}M^{[\beta\lambda]}]=\frac{3}{5}\nabla_\lambda^C[\sqrt{-g}M^{(\beta\lambda)}] \ .
\end{equation}
Using this result, (\ref{eq:con1s}) can be written as
\begin{equation} \label{eq:result-symmetric}
\nabla^C_\lambda\left[\sqrt{-g} \left(\delta_\alpha^\lambda M^{(\beta\nu)}-\frac{1}{5}\left\{\delta_\alpha^\nu M^{(\beta\lambda)}+\delta_\alpha^\beta M^{(\nu\lambda)}\right\}\right)\right] =0 \ .
\end{equation}
This equation can be formally solved in general for theories of the form $f(B,Q_S,Q_A)=\tilde{f}(B,Q_S)+\hat{f}(Q_A)$ by means of elementary algebraic manipulations. Key elements in the derivation are the following:
\begin{enumerate}
\item Denote ${P_\mu}^\nu\equiv B_{(\mu\alpha)}g^{\alpha \nu}$ and $F_{\mu\nu}\equiv B_{[\mu\nu]}=\frac{1}{2}\left[\partial_\mu C_\nu-\partial_\nu C_\mu\right]$  (because there is no torsion), where $C_\nu\equiv C^\lambda_{\nu\lambda}$. Raise one index in (\ref{eq:met1s}) with the metric, and bring all the $F_{\mu\nu}$ terms to the right-hand side to get
\begin{equation}\label{eq:quadratic}
\tilde{f}_B {P_\mu}^\nu+2\tilde{f}_{Q_S} {P_\mu}^\lambda {P_\lambda}^\nu
-\frac{\tilde{f}}{2}{\delta_\mu}^\nu=\kappa^2 {T_\mu}^\nu+{\tau_\mu}^\nu \ ,
\end{equation}
where
\begin{equation}\label{eq:tau}
{\tau_\mu}^\nu\equiv 2\left[\hat{f}_{Q_A} {F_\mu}^\lambda {F_\lambda}^\nu-\frac{\hat{f}}{4} {\delta_\mu}^\nu\right] \ .
\end{equation}
 Since $B\equiv {P_\mu}^\mu$ and $Q_S\equiv {[P^2]_\mu}^\mu$, (\ref{eq:quadratic}) can be seen as a quadratic matrix equation implying that  ${P_\alpha}^\beta$ is an algebraic function of the elements on the right-hand side of (\ref{eq:quadratic}), which can be seen as the matter sources. Note, in this sense, that  ${\tau_\mu}^\nu$ is formally identical to the stress-energy tensor of a non-linear theory of electrodynamics with $F_{\mu\nu}$ playing the role of the field strength tensor\footnote{Note in passing  that if we include a $Q_A^*$ piece in the Lagrangian density, a contribution $2\hat{f}_{Q_A^*}{F_\mu}^{\lambda} {F_\lambda}^{* \nu}$ to ${\tau_\mu}^{\nu}$ would also appear in (\ref{eq:tau}), thus completing the typical form of the stress-energy tensor of a non-linear theory of electrodynamics.}, and that it does not depend on $C^\lambda_{\alpha\beta}$.

\item  Decompose $M^{(\beta\nu)}=\tilde{f}_B g^{\beta\nu}+2\tilde{f}_{Q_S}B^{(\beta\nu)}$ as $M^{(\beta\nu)}=g^{\beta\alpha}
{\Sigma_\alpha}^\nu$, with
\begin{equation}\label{eq:Sigma}
{\Sigma_\alpha}^\nu=\tilde{f}_B  {\delta_\alpha}^\nu +2\tilde{f}_{Q_S}{P_\alpha}^\nu \ .
\end{equation}
Assume that ${\Sigma_\alpha}^\nu$ is invertible (which is true for models sufficiently close to GR). The invertibility of ${\Sigma_\alpha}^\nu$ implies that $M^{(\beta\nu)}$ is also invertible. Denote as $W_{\mu\alpha}\equiv {[\Sigma^{-1}]_\mu}^\kappa g_{\kappa \alpha}$ the inverse of $M^{(\beta\nu)}$, which is also a symmetric tensor.

\item Introduce the vectors $l_\alpha\equiv \partial_\alpha \ln \sqrt{-g}$  and $\sigma_\alpha\equiv \partial_\alpha \ln \sqrt{\det \Sigma}$. Note that  $l_{\alpha}$ is the gradient of a scalar function and that in GR it coincides with $C^\mu_{\mu\alpha}$. In the (more general) case considered here, we will see that $C^\mu_{\mu\alpha}$ may also pick up a pure vector contribution, i.e., it is not just given by the gradient of a scalar.  Now use the relation $W_{\beta\nu}\partial_\alpha M^{(\beta\nu)}=2(\sigma_\alpha-l_\alpha)$, which is analogous to $g_{\beta\nu}\partial_\alpha g^{\beta\nu}=-2l_\alpha$, to show that $\partial_\lambda M^{(\lambda\beta)}+C^\beta_{\lambda\rho}M^{(\lambda\rho)}=(4l_\lambda-5C_\lambda+5\sigma_\lambda) M^{(\lambda\beta)}$.

\item Expand (\ref{eq:result-symmetric}) and use the above notation and results to obtain
\begin{equation}\label{eq:C_apriori}
\partial_\alpha M^{(\beta\nu)}+C^\beta_{\alpha\lambda}M^{(\lambda\nu)}+
C^\nu_{\alpha\lambda}M^{(\lambda\beta)}=\Phi_\lambda({\delta_\alpha}^\nu  M^{(\beta\lambda)}+{\delta_\alpha}^\beta  M^{(\nu\lambda)})-\Omega_\alpha M^{(\beta\nu)} \ ,
\end{equation}
where $\Omega_\alpha\equiv (l_\alpha-C_\alpha)$, and $\Phi_\lambda\equiv \sigma_\lambda+\Omega_\lambda$.
\end{enumerate}
From (\ref{eq:C_apriori}) it is easy to show that
\begin{equation}\label{eq:Conn0}
C^\lambda_{\mu\nu}=\frac{M^{\lambda\rho}}{2}\left[\partial_\mu W_{\rho\nu}+\partial_\nu W_{\rho\mu}-\partial_\rho W_{\mu\nu}\right]-\frac{1}{2}\left[\delta^\lambda_\mu \Omega_\nu+\delta^\lambda_\nu \Omega_\mu-W_{\mu\nu}M^{(\lambda\rho)}(2\Phi_\rho+\Omega_\rho)\right] \ .
\end{equation}
Given the form of this solution, it is natural to introduce an auxiliary metric $h_{\mu\nu}$ and its inverse (this is just a choice of the conformal representative),
\begin{equation}\label{eq:hmn-def}
h_{\mu\nu}\equiv (\sqrt{\det \Sigma}) W_{\mu\nu} \ , \ h^{\mu\nu}=M^{(\mu\nu)}/\sqrt{\det \Sigma} \ ,
\end{equation}
in terms of which (\ref{eq:Conn0}) takes the form
\begin{equation}\label{eq:Conn1}
C^\lambda_{\mu\nu}=L^\lambda_{\mu\nu}(h)-\frac{1}{2}\left[\delta^\lambda_\mu \Phi_\nu+\delta^\lambda_\nu \Phi_\mu-3 h_{\mu\nu}h^{\lambda \rho}\Phi_\rho\right] \ ,
\end{equation}
where $L^\lambda_{\mu\nu}(h)$ represents the Christoffel symbols of $h_{\alpha\beta}$. With this notation, we also find that $\Phi_\lambda\equiv \partial_\lambda \ln \sqrt{-h}-C_\lambda$. Note that ${\Sigma_\alpha}^\nu$ in Eq.(\ref{eq:Sigma}), which is a function of the matter sources and the field $\Phi_\alpha$ (see the discussion following \ref{eq:quadratic}),  determines the relative deformation between the metrics $g_{\mu\nu}$ and $h_{\mu\nu}$. This deformation extends the idea of conformal and disformal transformations to a more general scenario. Similar deformation matrices have been considered in the literature in an heuristic manner (see \cite{Capozziello:2011et} for details and references), which contrasts with the naturalness of our approach. 

The above manipulations and definitions allow us to rewrite (\ref{eq:con1s}) in a more convenient form starting with its representation given in (\ref{eq:quadratic}).
It is easy to see that (\ref{eq:quadratic}) can be written as
\begin{equation}
{P_\mu}^\alpha {\Sigma_\alpha}^\nu = \kappa^2 {T_\mu}^\nu +{\tau_\mu}^\nu +\frac{\tilde{f}}{2}{\delta_\mu}^\nu  \ .
\end{equation}
Since ${P_\mu}^\alpha {\Sigma_\alpha}^\nu\equiv R_{(\mu\beta)}h^{\beta \nu}\sqrt{\det \Sigma} $, we finally get
\begin{equation}\label{eq:Rmn0}
R_{(\mu\alpha)} (C) =\left[\kappa^2 {T_\mu}^\nu +{\tau_\mu}^\nu +\frac{\tilde{f}}{2}{\delta_\mu}^\nu \right] \frac{h_{\nu\alpha}}{\sqrt{\det \Sigma}} \ .
\end{equation}
Using (\ref{eq:Conn1}) to construct $R_{(\mu\alpha)} (C)$, we find
\begin{equation}
R_{(\mu\alpha)} (C) \equiv  R_{\mu\alpha} (h)-\frac{3}{2}\Phi_\mu \Phi_\alpha  \ ,
\end{equation}
where $R_{\mu\alpha} (h)$ represents the Ricci tensor of the metric $h_{\mu\nu}$, which is symmetric by construction (because $L^\lambda_{\lambda \rho}(h)=\partial_\rho \ln \sqrt{-h}$ and, therefore, $R_{[\mu\nu]}=\frac{1}{2}\left[\partial_\mu L^\lambda_{\lambda \nu}(h)-\partial_\nu L^\lambda_{\lambda \mu}(h)\right]=0$) . This allows to express (\ref{eq:Rmn0}) as
\begin{equation}\label{eq:Rmn1}
{R_\mu}^\alpha (h) =\frac{1}{\sqrt{\det \Sigma}}\left[\kappa^2 {T_\mu}^\alpha +{\tau_\mu}^\alpha +\frac{\tilde{f}}{2}{\delta_\mu}^\alpha\right]  +\frac{3}{2}\Phi_\mu \Phi_\nu h^{\nu\alpha}\ .
\end{equation}
The above equation can also be written in terms of the metric $g_{\mu\nu}$ only. Since the difference between any two connections is a tensor, one finds that $0=\nabla^h_\lambda h_{\mu\nu}=\nabla^g_\lambda  h_{\mu\nu}-X^\alpha_{\lambda\mu}  h_{\alpha\nu}-X^\alpha_{\lambda\nu}  h_{\mu\alpha}$, where $X^\alpha_{\lambda\mu} $ is a tensor  and $\nabla^h_\lambda$ represents the covariant derivative associated to the Levi-Civita connection of the metric $h_{\alpha\beta}$, i.e., $\nabla^h_\lambda A_\mu=\partial_\lambda A_\mu -L^\alpha_{\lambda\mu}(h) A_\alpha$. Straightforward manipulations allow to obtain an explicit expression for $X^\alpha_{\lambda\mu}$, which becomes
\begin{equation}
X^\alpha_{\lambda\mu}=\frac{h^{\alpha\rho}}{2}\left[\nabla^g_\lambda h_{\rho\mu}+\nabla^g_\mu h_{\rho\lambda}-\nabla^g_\rho h_{\lambda\mu}\right] \ .
\end{equation}
With this expression, one can rewrite $R_{\mu\alpha} (h)$ as
\begin{equation}\label{eq:Riccih-to-Riccig}
R_{\nu\beta} (h)=R_{\nu\beta} (g)+\nabla_\lambda^L X^\lambda_{\nu\beta}-\nabla_\nu^L X^\lambda_{\lambda\beta}+X^\lambda_{\nu\beta}X^\alpha_{\alpha\lambda}-X^\lambda_{\alpha\beta}X^\alpha_{\nu\lambda}.
\end{equation}
In many cases of interest, however, it is possible and much more convenient to work directly with $h_{\mu\nu}$ \cite{Guendelman:2013sca, Olmo:2013mla, Lobo:2013adx,or12a1,or12a2,or12a3,Barragan:2010qb}.

The equations (\ref{eq:direct_coupling}) governing the behavior of the antisymmetric part of the Ricci tensor can be written as
\begin{equation}\label{eq:Proca_general}
\partial_\lambda[\sqrt{-g}\hat{f}_{Q_A}F^{\beta\lambda}] =\frac{3}{10}\sqrt{-h}\Phi_\rho h^{\rho\beta}=\frac{3}{10}\sqrt{-g}\Phi^\alpha{\Sigma_\alpha}^\beta   \ ,
\end{equation}
where $F^{\beta\lambda}=g^{\beta\alpha}g^{\lambda \rho}F_{\alpha\rho}$, and $F_{\alpha\rho}=\frac{1}{2}\left[\partial_\alpha C_\rho-\partial_\rho C_\alpha\right]=\frac{1}{2}\left[\partial_\alpha \Phi_\rho-\partial_\rho \Phi_\alpha\right]$, because the difference between $C_\rho$ and $\Phi_\rho$ is the gradient of a scalar.

If we particularize our general $f(B,Q_S,Q_A)$ Lagrangian to the simple case $f(B,Q_S,Q_A)=B+\alpha Q_A$, we recover a case studied in \cite{Buchdahl} (see also \cite{Vollick:2006uq, Vitagliano:2010pq}).  In this case $\hat{f}_{Q_A}=\alpha$, and from (\ref{eq:Sigma}) one finds that ${\Sigma_\alpha}^\nu={\delta_\alpha}^\nu$, which implies  $h_{\mu\nu}=g_{\mu\nu}$. Elementary manipulations allow to show that the theory is equivalent to the Einstein-Proca system, with $\Phi_\lambda\equiv \partial_\lambda \ln \sqrt{-h}-C^{\rho}_{\rho\lambda}$ playing the role of the (Proca) vector potential.

Our results allow to study generalizations of the Einstein-Proca system to the case of non-linear Proca Lagrangians, $\hat{f}_{Q_A}\neq $constant, and beyond, with ${\Sigma_\alpha}^\beta  \neq \delta_\alpha^\beta$. It is important to note that, regardless of the particular Lagrangian chosen, the vector and metric field equations are always second-order (though non-linear in general). This can be explicitly seen from   (\ref{eq:Proca_general}), which governs the vector field, and from  (\ref{eq:Rmn1}) together with the relations (\ref{eq:hmn-def}) and (\ref{eq:Riccih-to-Riccig}), which specify the algebraic relation between $h_{\mu\nu}$ and $g_{\mu\nu}$ and which do not contain higher than second-order derivatives of either of these metrics.

\subsection{Spaces with a non-symmetric connection \label{sec:post}}

We now consider a general family of Ricci-squared theories  but assume that the connection is made symmetric  {\it a posteriori}, i.e., torsion is allowed by construction but is set to zero after performing the variation of the action.

To proceed, we shall closely follow the derivation performed in \cite{olmo11b,olmo11b2}. Now the theory under analysis is $f(R,Q_S,Q_A)$ with $R\equiv g^{\mu\nu}R_{(\mu\nu)}$, $Q_S= R^{(\mu\nu)}R_{(\mu\nu)}$, and $Q_A=R^{[\mu\nu]}R_{[\mu\nu]}$, where we have replaced $B_{\mu\nu}$ tensors by the $R_{\mu\nu}$ ones, as compared to the previous subsection, to take into account that they include now a torsion contribution (for simplicity we shall neglect the $Q_A^*$ piece here as well). Thus $M^{\beta \nu}$ in (\ref{eq:def_M}) now becomes
\begin{equation}\label{eq:def_Mpost}
{P_\alpha}^{\beta\mu\nu} = \delta_\alpha^\mu\left[f_R g^{\beta\nu}+2f_{Q_S} R^{(\beta\nu)}+2f_{Q_A} R^{[\beta\nu]}\right]\equiv \delta_\alpha^\mu M^{\beta\nu} \ .
\end{equation}
%Recall that all  indices are raised with $g^{\mu\nu}$.
To work out in this case the equation for the connection, we insert (\ref{eq:def_Mpost}) in (\ref{eq:Gamn_a2}) and tracing over $\alpha$ and $\nu$ we obtain $\nabla_{\lambda}^C [\sqrt{-g}M^{\beta \lambda}]=(2\sqrt{-g}/3)[S_{\lambda \sigma}^{\sigma}M^{\beta \lambda} + (3/2) S_{\lambda \mu}^{\beta} M^{\lambda \mu}]$ and with this we can rewrite the connection equation (\ref{eq:Gamn_a2}) as
\begin{equation} \label{eq:con0}
\frac{1}{\sqrt{-g}}\nabla_{\alpha}^C [\sqrt{-g} M^{\beta \nu}]=S_{\alpha\lambda}^{\nu}M^{\beta \lambda}-S_{\beta \lambda}^{\nu} M^{\lambda \nu}-S_{\alpha \lambda}^{\lambda} M^{\beta \nu}+\frac{2}{3} \delta_{\alpha}^{\nu}S_{\lambda \sigma}^{\sigma}M^{\beta \lambda}
\end{equation}
whose symmetric and antisymmetric components are, respectively,
\begin{eqnarray}
\frac{1}{\sqrt{-g}}\nabla_{\alpha}^C [\sqrt{-g} M^{(\beta \nu)}]&=&S_{\alpha\lambda}^{\nu}M^{[\beta \lambda]}-S_{\beta \lambda}^{\nu} M^{[\lambda \nu]}-S_{\alpha \lambda}^{\lambda} M^{(\beta \nu)}+ \nonumber \\ &+&\frac{S_{\lambda \sigma}^{\sigma}}{3}(\delta_{\alpha}^{\nu} M^{\beta \lambda} + \delta_{\alpha}^{\beta} M^{\nu \lambda}) \label{eq:con1} \\
\frac{1}{\sqrt{-g}}\nabla_{\alpha}^C [\sqrt{-g} M^{[\beta \nu]}]&=&S_{\alpha\lambda}^{\nu}M^{(\beta \lambda)}-S_{\beta \lambda}^{\nu} M^{(\lambda \nu)}-S_{\alpha \lambda}^{\lambda} M^{[\beta \nu]}+ \nonumber \\ &+&\frac{S_{\lambda \sigma}^{\sigma}}{3}(\delta_{\alpha}^{\nu} M^{\beta \lambda} - \delta_{\alpha}^{\beta} M^{\nu \lambda}) \label{eq:con2}
\end{eqnarray}
Before setting the torsion to zero and comparing with the results of the previous subsection, we carry out a few additional manipulations to simplify the structure of these equations, which will bring new insights on the coupling between the symmetric and antisymmetric parts of $M^{\alpha\beta}$.
We thus introduce a new connection $\widetilde{\Gamma}_{\alpha \beta}^{\lambda}$ such that
\begin{equation}
\Gamma_{\alpha \beta}^{\lambda}=\widetilde{\Gamma}_{\alpha \beta}^{\lambda} - \frac{2}{3} \delta_{\nu}^{\lambda} S_{\sigma \mu}^{\sigma}.
\end{equation}
This change implies that $\widetilde{S}_{\mu\nu}^{\lambda} \equiv \widetilde{\Gamma}_{[\mu\nu]}^{\lambda}$ satisfies $\widetilde{S}_{\sigma \nu}^{\sigma}=0$. The symmetric and antisymmetric components of $\widetilde{\Gamma}_{\mu\nu}^{\lambda}$ are related to those of $\Gamma_{\mu\nu}^{\lambda}$ by
\begin{eqnarray}
C_{\mu\nu}^{\lambda}&=&\widetilde{C}_{\mu\nu}^{\lambda}-\frac{1}{3} (\delta_{\nu}^{\lambda}S_{\mu}+\delta_{\mu}^{\lambda} S_{\nu}) \label{eq:newconsym} \\
S_{\mu\nu}^{\lambda}&=&\widetilde{S}_{\mu\nu}^{\lambda}-\frac{1}{3} (\delta_{\nu}^{\lambda}S_{\mu}-\delta_{\mu}^{\lambda} S_{\nu}) \label{eq:newconsym2},
\end{eqnarray}
where $S_{\mu}=S_{\lambda \mu}^{\lambda}$, which is in general nonvanishing, as opposed to $\widetilde{S}_{\mu}=\widetilde{S}_{\lambda \mu}^{\lambda}=0$, as easily seen from (\ref{eq:newconsym2}). The relation of the Ricci tensor in the connections $\Gamma_{\alpha \beta}^{\lambda}$ and $\widetilde{C}_{\alpha \beta}^{\lambda}$ becomes (from now on we denote $\widetilde{\nabla}_{\mu} \equiv \nabla^{\widetilde{C}}_{\mu}$)
\begin{equation}
R_{\beta \nu}(\Gamma)=R_{\beta \nu}(\widetilde{C})+\frac{1}{3}[\widetilde{\nabla}_{\nu} S_{\beta}-\widetilde{\nabla}_{\beta}S_{\nu}]+\widetilde{\nabla}_{\lambda}\widetilde{S}_{\nu\beta}^{\lambda}-\widetilde{S}_{\kappa \beta}^{\lambda}\widetilde{S}_{\nu\lambda}^{\kappa},
\end{equation}
which in terms of the symmetric and antisymmetric components reads
\begin{eqnarray}
R_{(\beta \nu)} (\Gamma)&=&R_{(\beta \nu)} (\widetilde{C}) -\widetilde{S}_{\kappa \beta}^{\lambda} \widetilde{S}_{\nu\lambda}^{\kappa} \label{eq:R1a}\\
R_{[\beta \nu]}(\Gamma)&=&R_{[\beta \nu]}(\widetilde{C})+\frac{1}{3}[\widetilde{\nabla}_{\nu} S_{\beta}-\widetilde{\nabla}_{\beta}S_{\nu}]+\widetilde{\nabla}_{\lambda}\widetilde{S}_{\nu\beta}^{\lambda}. \label{eq:R1b}
\end{eqnarray}
Let us recall that from (\ref{Ricci-antisymmetric}) and (\ref{eq:newconsym}) we have
\begin{equation} \label{eq:antiRiccicon2}
R_{[\beta \nu]}(\widetilde{C})=\frac{1}{2}[\widetilde{\nabla}_{\beta}\widetilde{C}_{\nu}-\widetilde{\nabla}_{\nu}\widetilde{C}_{\beta}],
\end{equation}
where $\widetilde{C}_{\beta}=\widetilde{C}_{\lambda \beta}^{\lambda}$. This equation simply tells us that in the connection $\widetilde{C}_{\mu\nu}^{\lambda}$ the antisymmetric part of the Ricci tensor can be expressed as the rotational of the vector $\widetilde{C}_{\mu}$.

Collecting all these results, the symmetric and antisymmetric parts of the connection equation (\ref{eq:con0}) become
\begin{eqnarray}
\frac{1}{\sqrt{-g}}\widetilde{\nabla}_{\alpha}[\sqrt{-g}M^{(\beta \nu)}]&=&{\Lambda^{(+)} }_{\alpha}^{\beta \nu \kappa \rho} M_{[\kappa \rho]} \label{eq:con1a} \\
\frac{1}{\sqrt{-g}}\widetilde{\nabla}_{\alpha}[\sqrt{-g}M^{[\beta \nu]}]&=&{\Lambda^{(-)} }_{\alpha}^{\beta \nu \kappa \rho} M_{(\kappa \rho)}, \label{eq:con1b}
\end{eqnarray}
where $M^{(\beta \nu)}=f_R g^{\beta \nu} + 2f_{Q_S} R^{(\beta \nu)}(\Gamma)$ and $M^{[\beta \nu]}=2f_{Q_A} R^{[\beta \nu]}(\Gamma)$ and we have defined the objects
\begin{equation}
{\Lambda^{(\pm)} }_{\alpha}^{\beta \nu \kappa \rho}=\left[\widetilde{S}_{\alpha \lambda}^{\nu} g^{\beta \kappa} \pm \widetilde{S}_{\alpha \lambda}^{\beta} g^{\nu \kappa} \right] g^{\lambda \rho}.
\end{equation}
These equations clearly show that the symmetric and antisymmetric parts of $M^{\beta \nu}$ are not directly coupled, which contrasts with  (\ref{eq:direct_coupling}). Instead, they couple to each other via torsion, through the traceless tensor $\widetilde{S}_{\alpha \lambda}^{\nu}$.\\

\subsubsection{Torsionless scenario}

In general, when the  tensor $\widetilde{S}_{\alpha \lambda}^{\nu}$ vanishes, we find that ${\Lambda^{(\pm)} }_{\alpha}^{\beta \nu \kappa \rho}=0$, which implies the decoupling between  $M^{(\beta \nu)}$ and $M^{[\beta \nu]}$. This choice still leaves unspecified a part of the torsion tensor,  $S_{\alpha \lambda}^{\nu}=\frac{1}{3}(\delta_{\alpha}^{\nu}S_{\lambda}-\delta_{\lambda}^{\nu} S_{\alpha})$, which is determined by a vector and needs not be zero.  From (\ref{eq:R1a}) and (\ref{eq:R1b}) we see that in this case
\begin{eqnarray}\label{eq:G2C}
R_{(\beta \nu)}(\Gamma)&=&R_{(\beta \nu)}(\widetilde{C}) \\
R_{[\beta \nu]}(\Gamma)&=&R_{[\beta \nu]}(\widetilde{C})+\frac{1}{3}[\widetilde{\nabla}_{\nu} S_{\beta}-\widetilde{\nabla}_{\beta}S_{\nu}]\nonumber \\ &=&\frac{1}{2}[\widetilde{\nabla}_{\nu} \Psi_{\beta}-\widetilde{\nabla}_{\beta}\Psi_{\nu}] =\frac{1}{2}[{\partial}_{\nu} \Psi_{\beta}-{\partial}_{\beta}\Psi_{\nu}] \ , \label{eq:FmnPost}
\end{eqnarray}
 where $\Psi_\nu\equiv \widetilde{C}_\nu-\frac{2}{3}S_\nu$%=C_\nu+S_\nu$
, and the last result follows because $ \widetilde{C}^\lambda_{\mu\nu}$ is a symmetric connection.

Since the connection does not appear explicitly in the expression of $F_{\beta \nu}\equiv R_{[\beta \nu]}(\Gamma)$, the metric variation of these theories can be manipulated in the same way as in the previous subsection. Starting with the analogous of (\ref{eq:met1s}),
\begin{equation}
f_R R_{(\mu\nu)}+2g^{\alpha\beta}\left[f_{Q_S}R_{(\mu\alpha)}R_{(\nu\beta)}+f_{Q_A}F_{\mu\alpha}F_{\nu\beta}\right]-\frac{f}{2}g_{\mu\nu} = \kappa^2 T_{\mu\nu} \ , \label{eq:met1sPost}
\end{equation}
the $F_{\beta \nu}$-dependent terms can be transferred to the right-hand side of the field equations (we may assume, as before, that the Lagrangian can be written as $f(R,Q_S,Q_A)=\tilde{f}(R,Q_S)+\hat{f}(Q_A)$, though this aspect is not essential for what follows). Then one index  is raised with $g^{\nu\alpha}$ and the resulting algebraic quadratic equation for the matrix ${P_\mu}^\nu$ can be (formally) solved by means of algebraic manipulations, which implies that the tensor ${\Sigma_\alpha}^\mu$ defined in (\ref{eq:Sigma}) (with the notational replacement $B\to R$) is just a function of the matter sources and the field $F_{\beta \nu}$. Using the definitions
\begin{equation}\label{eq:hmn}
h^{\mu\nu}=g^{\mu\alpha}{\Sigma_\alpha}^\nu/\sqrt{\det \Sigma}  \ , \  h_{\mu\nu}\equiv (\sqrt{\det \Sigma}) {[\Sigma^{-1}]_\mu}^\alpha g_{\alpha\nu} \ ,
\end{equation}
and taking into account (\ref{eq:G2C}), we find that (\ref{eq:met1sPost}) can be expressed as
\begin{equation}\label{eq:Rmn0Post}
R_{(\mu\alpha)} (\widetilde{C}) =\left[\kappa^2 {T_\mu}^\nu +{\tau_\mu}^\nu +\frac{\tilde{f}}{2}{\delta_\mu}^\nu \right] \frac{h_{\nu\alpha}}{\sqrt{\det \Sigma}} \ ,
\end{equation}
where ${\tau_\mu}^\nu$ is defined exactly as in (\ref{eq:tau}).
On the other hand, (\ref{eq:con1a}) and  (\ref{eq:con1b}) turn into
\begin{eqnarray}
\widetilde{\nabla}_{\alpha}[\sqrt{-h}h^{\beta \nu}]&=&0 \label{eq:con2a} \\
\widetilde{\nabla}_{\alpha}[\sqrt{-g}\hat{f}_{Q_A}F^{[\beta \nu]}]&=&0 \ . \label{eq:con2b}
\end{eqnarray}
It is easy to verify that (\ref{eq:con2a}) implies that $\widetilde{C}^\lambda_{\mu\nu}$ is given by the Christoffel symbols of $h_{\beta \nu}$, i.e., $\widetilde{C}^\lambda_{\mu\nu}=L^\lambda_{\mu\nu}(h)$. As a result, the connection $\Gamma^\lambda_{\mu\nu}$ can be written as
\begin{equation}\label{eq:Conn1post}
\Gamma^\lambda_{\mu\nu}=L^\lambda_{\mu\nu}(h)-\frac{2}{3}\delta^\lambda_\nu S_\mu \ .
\end{equation}
The fact that $\widetilde{C}^\lambda_{\mu\nu}$ is metric compatible, implies that $R_{[\beta\nu]}(\widetilde{C})=0$ because $\widetilde{C}_\mu\equiv \widetilde{C}^\lambda_{\lambda\mu}=\partial_\mu \ln \sqrt{-h}$. A direct consequence of this is that  $F_{\mu\nu}\equiv \frac{1}{2}[{\partial}_{\nu} \Psi_{\beta}-{\partial}_{\beta}\Psi_{\nu}]=- \frac{1}{3}[{\partial}_{\nu} S_{\beta}-{\partial}_{\beta}S_{\nu}]$. This means that in the completely torsionless case  $F_{\mu\nu}\equiv 0$, i.e., the Ricci tensor of the connection is totally symmetric, and the antisymmetric part plays no role in the dynamics (note that in that case ${\tau_\mu}^\nu\equiv0$). The field equations for these torsionless theories, therefore, boil down to the following
\begin{equation}\label{eq:Rmn0Post}
{R_\mu}^\nu (h) =\frac{1}{\sqrt{\det \Sigma}}\left[\kappa^2 {T_\mu}^\nu +\frac{\tilde{f}}{2}{\delta_\mu}^\nu \right]  \ ,
\end{equation}
and should be compared with Eqs. (\ref{eq:Rmn1}) and  (\ref{eq:Proca_general}) corresponding to the {\it a priori} torsionless case. Eq. (\ref{eq:Rmn0Post}) must be supplemented with the field equations for the matter sources. Examples with perfect fluids, scalar fields, and electromagnetic fields have been considered in \cite{Lobo:2013adx,Olmo:2011sw,Barragan:2010qb,or12a1,or12a2,or12a3}.

Before concluding this section, we note that (\ref{eq:met1sPost}) can be written as (recall that ${P_\mu}^\nu\equiv R_{(\mu\alpha)}g^{\alpha \nu}$)
\begin{equation}\label{eq:quadratic2}
2f_{Q_S}\left(\hat{P}+\frac{{f_R}}{4f_{Q_S}}\hat{I}\right)^2=\left(\frac{f_R^2}{8f_{Q_S}}+\frac{f}{2}+\kappa^2\hat{T}\right)\hat{I} \ ,
\end{equation}
where the hat represents a matrix. In vacuum, $\hat{T}=0$, implies that $\hat{P}$ can be written as $\hat{P}=\Lambda(R^{vac},Q^{vac}_S)\hat{I}$, where the explicit form of $\Lambda(R^{vac},Q^{vac}_S)$ can be found straightforwardly from (\ref{eq:quadratic2}) but is unessential for the current discussion. The point is that the relations  $R^{vac}={P_{\mu}}^{\mu}=4\Lambda(R^{vac},Q^{vac}_S)$, and $Q_S^{vac}={[P^2]_{\mu}}^{\mu}=4\Lambda^2(R^{vac},Q^{vac}_S)$ imply that the values of $R^{vac}$ and $Q^{vac}_S$ are constant and in the relation $Q^{vac}_S=(R^{vac})^2/4$. In addition, $\hat{P}=\Lambda(R^{vac},Q^{vac}_S)\hat{I}$ also implies that $h_{\mu\nu}$ and $g_{\mu\nu}$ are related by a constant conformal factor [see Eqs. (\ref{eq:Sigma}) and (\ref{eq:hmn})]. As a result, (\ref{eq:Rmn0Post}) tells us that $R_{\mu\nu}(h)=C^{vac}h_{\mu\nu} \ \leftrightarrow \ R_{\mu\nu}(g)=\tilde{C}^{vac}g_{\mu\nu} $, with $C^{vac}$ and $\tilde{C}^{vac}$ constant (and identical in an appropriate system of units), which confirms that the vacuum equations coincide with the vacuum Einstein equations with a cosmological constant (whose  magnitude depends on the particular gravity Lagrangian). This shows that Palatini  $f(R,Q_S,Q_A)$ theories with vanishing torsion {\it a posteriori} recover the usual Einstein-de Sitter equations in vacuum. These theories, therefore, do not introduce any new propagating degrees of freedom besides the standard massless spin-2 gravitons and are free from the ghost-like instabilities present in the (higher-derivative) metric formulation of these same theories.

\section{Connections with an invariant volume} \label{sec:IV}

In the previous section we have seen that in theories with a symmetric connection (regardless of whether this condition is imposed {\it a priori} or {\it a posteriori}) the existence of an auxiliary metric $h_{\mu\nu}$ capturing an important part of the connection is always present. In this section we comment on this important aspect to explain the reasons for its existence and role.

The conditions for a symmetric connection to be Riemannian, i.e., to be the Levi-Civita connection of a metric, have been studied in the literature \cite{SymConn1,SymConn2,SymConn3,SymConn4,SymConn5}. These conditions are intimately related to the existence of a conserved volume, i.e., of a scalar density such that $\nabla_\alpha \sqrt{-h}=0$.  It seems to be well established that a necessary and sufficient condition for a symmetric connection to have an invariant volume is that $R^\lambda_{\lambda \mu\nu}=0$ which, according to our definition (\ref{eq:Riemann}), is equivalent to $R_{[\mu\nu]}=0$.   This condition can be understood as follows. If for a given symmetric connection $C^{\alpha}_{\mu\nu}$ there exists a rank-two, non-degenerate tensor $h_{\alpha\beta}$ such that $\nabla^C_\alpha \sqrt{-h}=0$, one has that $\nabla^C_\alpha \sqrt{-h}=\partial_\alpha \sqrt{-h}-C^\lambda_{\lambda\alpha}\sqrt{-h}=0$ or, equivalently, that $C_\alpha\equiv C^\lambda_{\lambda\alpha}=\partial_\alpha \ln\sqrt{-h}$. Since $C_\alpha$ is the gradient of a scalar, it follows from (\ref{Ricci-antisymmetric}) that $R_{[\mu\nu]}(C)=0$. One should note, in this sense, that for a given $h_{\alpha\beta}$ any two connections satisfying $\nabla^C_\alpha \sqrt{-h}=0=\nabla^{\widehat{C}}_\alpha \sqrt{-h}$ must be related by
\begin{equation}\label{eq:InvVol}
C^{\alpha}_{\mu\nu}=\widehat{C}^{\alpha}_{\mu\nu}+\frac{1}{5}\left(\delta^\alpha_\mu\psi_\nu+\delta^\alpha_\nu\psi_\mu\right)-h_{\mu\nu}\psi^\alpha \ ,
\end{equation}
where $\psi^\alpha\equiv h^{\alpha\lambda}\psi_\lambda$. As can be readily verified, this is so because $C^\lambda_{\lambda\mu}=\widehat{C}^\lambda_{\lambda\mu}$. Therefore, connections related in this way possess the same invariant volume.

When $R_{[\mu\nu]}(C)\neq 0$, it is possible to introduce a new (symmetric) connection $\widetilde{C}_{\alpha\beta}^{\lambda}$ such that $R_{[\beta \nu]}(\widetilde{C})=0$.  The connection $\widetilde{C}_{\alpha\beta}^{\lambda}$ is thus associated with a rank-two tensor. This is achieved by defining
\begin{equation}
{C}_{\alpha \beta}^{\lambda}=\widetilde{C}_{\alpha\beta}^{\lambda}+\delta_{\beta}^{\lambda}\varphi_{\alpha}+\delta_{\alpha}^{\lambda}\varphi_{\beta} ,\label{eq:newconn}
\end{equation}
where $\varphi_{\mu}$ is a vector field. From (\ref{eq:newconn}) we find that
\begin{equation}
R_{[\beta \nu]}({C})=R_{[\beta \nu]}(\widetilde{C})+\frac{5}{2}(\widetilde{\nabla}_{\beta} \varphi_{\nu}-\widetilde{\nabla}_{\nu}\varphi_{\beta}) \label{eq:con3c} \ ,
\end{equation}
and comparing with (\ref{eq:antiRiccicon2}) it is evident that in order to have $R_{[\beta \nu]}(\widetilde{C})=0$ we must take $\varphi_{\nu}=\frac{1}{5} {C}_{\nu}$.

We are now ready to understand the form of the connection found in Sec. \ref{sec:apriori} [see Eq.  (\ref{eq:Conn1})]. Since for that family of theories with an {\it a priori} symmetric connection $B_{[\mu\nu]}(C)\equiv R_{[\mu\nu]}(C)\neq 0$ in general, it is possible to introduce another symmetric connection $\widetilde{C}_{\alpha\beta}^{\lambda}$  such that $B_{[\mu\nu]}(\widetilde{C})=0$. This connection must be related to ${C}_{\alpha\beta}^{\lambda}$ through a transformation of the form (\ref{eq:newconn}). However, the last term in (\ref{eq:Conn1}) prevents a simple and direct identification of the vector $\varphi_{\beta}$ involved in the transformation. We must, therefore, take care  of that term first by considering another symmetric connection  of the form (\ref{eq:InvVol}), $\widehat{C}_{\alpha\beta}^{\lambda}$, which leaves invariant the same volume. We thus see that the identification $\psi_\lambda=-\frac{3}{2}\Phi_\lambda$ leads us to $\varphi_\lambda=-\frac{1}{5}\Phi_\lambda$, with $\widetilde{C}_{\alpha\beta}^{\lambda}=L_{\alpha\beta}^{\lambda}(h)$.

The case of Sec. \ref{sec:post} is much more direct, because when the connection is made symmetric by setting the torsion to zero, we find $R_{[\mu\nu]}(C)=0$ and the existence of an invariant volume is automatically guaranteed, as it follows from (\ref{eq:FmnPost}) and (\ref{eq:con2a}).

We would like to mention here that the existence of invariant volumes at each space-time point associated to the connection might be of physical relevance in high-energy scenarios. In fact, it has been suggested that such structures could be a manifestation, in the classical continuum limit, of a more fundamental discrete  quantum structure of space-time itself \cite{Bradonjic:2011jw,Anderson:1971pn}.
The crucial role played by the connection in the definition of such invariant volumes suggests that the {\it cellular} microscopic structure of space-time could be more closely related to the affine properties of the manifold than to its chrono-geometric properties. The consideration of independent metric and affine structures in the study of quantum gravity phenomenology \cite{Olmo:2011sw}  is thus an aspect that should be further explored.

\section{Summary and conclusions} \label{sec:V}

In this work we have studied the field equations of a rather general family of theories in which the gravity Lagrangian is a functional of the metric and an independent affine connection. We have shown that assuming the connection to be symmetric i) {\it a priori} or  ii) {\it a posteriori} has a non-trivial impact on the resulting field equations, which are different in general. For concreteness and to make contact with existing literature, we have particularized our analysis to Ricci-squared theories of the form $f(R, Q_S, Q_A)$, where $R=g^{\mu\nu}R_{\mu\nu}$, $Q_S=R_{(\mu\nu)}R^{(\mu\nu)}$, and $Q_A=R_{[\mu\nu]}R^{[\mu\nu]}$. We have been able to exactly solve for the connection $\Gamma^\lambda_{\mu\nu}$ in both cases, finding that it can be decomposed into a Levi-Civita part plus other terms determined by a vector degree of freedom [see Eqs.(\ref{eq:Conn1}) and (\ref{eq:Conn1post})].
 The Levi-Civita part is due to the existence of an invariant volume associated with an auxiliary metric $h_{\mu\nu}$,
which is related with $g_{\mu\nu}$ via a deformation matrix ${\Sigma_\alpha}^\beta$ that depends algebraically on the matter sources [see (\ref{eq:Sigma})]. In case i), the vector field is related to the contraction $C^\lambda_{\lambda\rho}$ of the (symmetric) connection plus the gradient of a scalar function, $\Phi_\rho\equiv \partial_\rho \ln \sqrt{-h}-C^\lambda_{\lambda\rho}$,  whereas in case ii) it is related to the contraction $S^\lambda_{\lambda\rho}$ of the (antisymmetric) torsion tensor plus a gradient, $\Psi_\nu\equiv \partial_\nu\ln\sqrt{-h}-\frac{2}{3}S_\nu$. Since this vector is the origin of the antisymmetric part of the Ricci tensor, case i) introduces, in general, a dynamical vector degree of freedom governed by (\ref{eq:Proca_general}). In case ii), however, setting the torsion to zero implies the vanishing of $S_\nu$ and hence the vanishing of $R_{[\mu\nu]}$. As a result, in case ii) the antisymmetric part of the Ricci tensor plays no role in the dynamics when the connection is made symmetric, which provides a solid justification for the choice $R_{[\mu\nu]}=0$ made in previous works \cite{or12a1,or12a2,or12a3,Olmo:2011sw,Barragan:2010qb,Olmo:2009xy}. In case i), the non-vanishing $R_{[\mu\nu]}$ implies the existence of a dynamical vector field, which allows to generalize the well known Einstein-Proca system to the case of non-linear Proca Lagrangians and beyond (see Sec. \ref{sec:apriori}).

To conclude, we remark that two important lessons of general interest follow from our analysis: 1) that if non-linear curvature corrections appear in the theory, then the consideration or not of torsion is crucial to correctly define the theory and its field equations, and 2) that (a part of) the connection in metric-affine theories with non-linear curvature corrections has associated an invariant volume which, in general, does not coincide with that defined by the metric appearing in the action.
A detailed study of the role of torsion in these theories is currently underway. The potential relation between the existence of an invariant volume and the lack of higher-order derivatives and ghosts in these theories will also be explored elsewhere.

\section*{Acknowledgments}

The work of G. J. O. has been supported by the Spanish grant FIS2011-29813-C02-02 and the JAE-doc program of the Spanish Research Council (CSIC). D.R.-G. is finantially supported by CNPq (Brazilian agency) through grant 561069/2010-7, and thanks the theoretical physics group at Valencia U., where part of this work was carried out, for hospitaly and partial support. The authors thank L. Fatibene and M. Francaviglia for useful discussions.

\end{document}